\documentclass[english,aps,rpm,reprint,onecolumn]{revtex4-2}
\usepackage[T1]{fontenc}
\usepackage[latin9]{inputenc}
\usepackage{geometry}
\usepackage{babel}
\usepackage{units}
\usepackage{amsmath}
\usepackage{amsthm}
\usepackage{amssymb}
\usepackage[unicode=true,pdfusetitle,
 bookmarks=true,bookmarksnumbered=false,bookmarksopen=false,
 breaklinks=false,pdfborder={0 0 1},backref=false,colorlinks=false]
 {hyperref}

\makeatletter
\numberwithin{equation}{section}
\numberwithin{figure}{section}
\theoremstyle{remark}
\newtheorem*{rem*}{\protect\remarkname}

\makeatother

\providecommand{\remarkname}{Remark}

\begin{document}
\title{The trouble with recording devices}
\author{Eric Tesse}
\email{eric.s.tesse@gmail.com}

\begin{abstract}
Quantum theory encounters a difficulty when attempting to describe
recording devices. If the recording is of events in which quantum
uncertainty plays a role, such as an experiment on a quantum system,
quantum theory is unable to correctly predict the probabilities of
both future and past states of the recording. The nature of this difficulty
will be laid out at the outset. A resolution then will be presented,
in which the Born rule will be lightly amended so as to correctly
predict all probabilities. The resolution will have the further benefit
of clarifying how quantum theory applies to an array of situations
in which the theory can be ambiguous, such as the descriptions of
continuous measurements, and of closed systems containing all observers.
\end{abstract}
\maketitle

\section{Quantum recording devices}

Quantum mechanics, as it is generally presented, appears to have difficulty
describing recording devices and the recordings that they create.

The Born rule states that the probability that $Y$ will be true at
time $t_{1}$, given that $X$ is true at $t_{0}$, is $P(Y;t_{1}|X;t_{0})=\frac{1}{Tr\left(\mathbb{P}_{X}(t_{0})\right)}Tr\left(\mathbb{P}_{X}(t_{0})\mathbb{P}_{Y}(t_{1})\right)$.
If Bayes' rule holds, then the condition can be switched: If $Y$
is the case at time $t_{1}$, the probability that $X$ had been the
case is $P(X;t_{0}|Y;t_{1})=\frac{1}{Tr\left(\mathbb{P}_{Y}(t_{1})\right)}Tr\left(\mathbb{P}_{X}(t_{0})\mathbb{P}_{Y}(t_{1})\right)$.
It will now be shown that when a system includes a recording devices,
there are cases in which these two equations can not both hold.

Consider the case a video camera capturing events unfolding. Let's
assume that the recording at $t_{0}$ is of the set-up of an experiment,
including the preparation of a measurement that is to be performed.
The recording at $t_{1}$ continues this recording through the determination
of the outcome of a measurement. Take $X$ at $t_{0}$ to be all states
of the experimental set-up as a whole in which the recording device
is actively recording, and has recorded the experiment being setup.
Take $Y$ at $t_{1}$ to be a set of states in which the recording
device has recorded the set-up as well as the measurement outcome
of an experimental run. We will assume that, due to the nature of
the preparation, the microstate at $t_{0}$ of the system to be measured
does not predetermine what the outcome will be.

Under these conditions, it must be the case that $P(X;t_{0}|Y;t_{1})=1$,
and for any microstate, $x$, in the $X$ subspace $P(Y;t_{1}|x;t_{0})<1$.
The first relation ensures that the final state of the recording determines
that the recording must have been in state $X$ at time $t_{0}$;
by the nature of a recording, this must hold. The second relation
states that whatever the outcome might be, it is not predetermined
by the initial microstate. The Born rule can not generate both of
these relations.

Take $\mathbb{P}_{X}(t_{0})$ to be the projection operator onto the
states of the experimental set-up as a whole in which the recording
device's state is an element of $X$. Similarly, take $\mathbb{P}_{Y}(t_{1})$
to be the projection operator onto the states of the experimental
set-up as a whole in which the recording device's state is an element
of $Y$.

By the Born rule, for $P(X;t_{0}|Y;t_{1})=1$ to be true, we must
have $Tr\left(\mathbb{P}_{X}(t_{0})\mathbb{P}_{Y}(t_{1})\right)=Tr\left(\mathbb{P}_{Y}(t_{1})\right)$.
For this to be the case, $\mathbb{P}_{Y}(t_{1})$ must project onto
a subspace of $\mathbb{P}_{X}(t_{0})$. Take this subspace at time
$t_{0}$ to be $\mathbb{P}_{Z}(t_{0})$. According to the Born rule,
if the experiment were to start within this subspace, the probability
that $Y$ will be the outcome of the experiment is $1$. However,
we assumed that for this experiment the initial microstate did not
predetermine the outcome. So there can be no such subspace of $X$. 

The Born rule therefore can not yield correct probabilities for states
of a recording, when the recording is of a sequence events in which
quantum uncertainty renders the sequence uncertain from the outset.
Note that there's nothing in the Born rule that says that it can not
be applied to the states of a recording device. It ought to be applicable.

There is no reason to believe that this represents a deep flaw in
quantum theory. It appears to be, and is, a technical difficulty.

The nature of the difficulty can be made plain if we consider a particular
example of the above scenario. For the experiment being recorded,
imagine a source of particles that produces positively charged spin
$\nicefrac{1}{2}$ fermions with momentum in the z-direction. A Stern-Gerlach
device is placed in the path of these particles, with the SG device's
magnetic field pointing in the y-direction. Particles deflected by
the field in the negative y direction are blocked, while those deflected
in the positive y-direction are detected (in a way that doesn't disturb
the spin or impede the particle). Beyond that is a second SG set-up,
with magnetic field aligned in x-direction, and beyond the 2nd SG
set-up is another detector; this one detects the direction that the
particle was deflected. At $t_{0}$ the recording has recorded the
set-up and run of the experiment through the time that the first detector
is triggered. At $t_{1}$ the recording has recorded through the outcome
seen in second detector.

Here, the difficulty is clear. The recording at $t_{0}$ implies that
the particle spin to be in the y-direction, while the recording at
$t_{1}$ implies that the particle spin at $t_{1}$ is in the x-direction.
Because the recording at $t_{1}$ implies the recording at $t_{0}$,
the recording at $t_{1}$ also implies that at $t_{0}$ the particle
spin was in the y-direction. However, if we propagate the states at
$t_{1}$ back to $t_{0}$, the particle spin will continue to be in
x-direction, rather than the y-direction. Due to the mismatched spins,
$\frac{1}{Tr\left(\mathbb{P}_{Y}(t_{1})\right)}Tr\left(\mathbb{P}_{X}(t_{0})\mathbb{P}_{Y}(t_{1})\right)\neq1$.

For the Hilbert space to be complete, it must include the vector corresponding
to the state of the experiment at time $t_{1}$, propagated back in
time to $t_{0}$. This is a vector with the first SG set-up is in
a state that measures a particle with spin in the y-direction, while
the particle's spin state is actually in the x-direction. However,
in all applications of quantum theory to SG measurements, it is recognized
that such a state is not physically possible. If a particle has been
measured with its spin in the y-direction, then at that moment in
time its spin must be in the y-direction. The difficulty arises because
this not been included as a formal aspect of the theory. (Because
the above thought experiment will prove useful in later discussions,
it will from here on out be referred to as our ``reference experiment''.)

The problem is a little unusual in that it is not in the least paradoxical.
Our intuitive reasoning is correct. Some element of our intuitive
reasoning simply appears to be missing from the theory.

It only remains to formalize our intuitive understanding. This formalization
will yield interesting benefits. A number of aspects of quantum theory
that are hazy because they are outside the confines of how the theory
is normally applied will be clarified. For example, it will allow
quantum theory to provide straightforward descriptions closed system
that include all observers. Because the description will rest solely
on the Born rule, this clarification will not rely on any interpretation
of quantum theory, it will be a feature of the theory itself. The
formulation will also yield straightforward descriptions of types
of measurements that sometimes cause consternation, such continuous
measurements, and measurements in which the measurement time is not
precisely determined. It will also eliminate as nonphysical several
scenarios that can raise paradoxes.

This will not have any impact on quantum interpretations. All valid
interpretations of quantum theory will remain valid. None will emerge
preferred, and none will be discarded.

(A wholly unfortunate, and hopefully unnecessary, parenthetical remark:
None of this has anything to do with wave-packet collapse. If you
propagate a particle through a SG set-up, its position always ends
up being correlated to its spin. In our reference experiment, you
can't end up with the portion of the wavefunction that has been deflected
in the y-direction having its spin in the x-direction. Nor has it
been assumed that the wavefunction collapses at time $t_{1}$. No
assumption about the nature of the wavefunction at time $t_{1}$ has
been made. We are free to assume that the recording has only captured
a portion of the wavefunction, and in a different portion, the recording
state captures a different outcome. We have simply applied the Born
rule as it is always applied. Within this article, we will only be
interested in how quantum theory can be applied to scientific observations
of our world. Any valid quantum interpretation ought to be consistent
with all such observations.)

\section{The physical subspace}

\subsection{\label{subsec:Introducing-the-physical}Introducing the physical
subspace}

There are physical systems whose internal states can reflect the state
of the outside world. Our brains are a prominent example. We sense
the outside world, and through our senses the state of the world is
registered in our brain states. The capability of physical systems
to register the state of the outside world within their internal states
is a requirement for performing scientific observations.

In classical systems, the ability of a system's state to mirror the
state of the outside world is due to fields. Particle states can be
mixed and matched in any fashion, and so the state of some set of
particles can not reveal anything about the states of other particles.
But field states in one region of space tell us a great deal about
the field state in other regions, and those fields also carry a great
deal of information about particle states, due to fields' interactions
with particles. It may be assumed for quantum systems as well, fields
play a central role in information gathering. But the physical mechanism
behind informational ties between subsystem states will not concern
us here. Here, we will simply use the tools of quantum theory to ensure
that such ties are respected.

The subspace of Hilbert state vectors that respect these ties between
subsystem states will be called the \emph{physical subspace}. As an
example of the physical subspace, if a detector detects a particle,
then in the physical subspace the particle must be within the detector.
In the full Hilbert space, detector states and particle states can
be mixed and matched in any fashion. The projection operator onto
the physical subspace will be represented by $\mathcal{P}$. (For
ease of notation, $\mathcal{P}$ will also sometimes be used to represent
the physical subspace itself, as a subspace and the projection operator
onto the subspace each uniquely specify the other.)

As a further illustration of the physical subspace, let's once again
turn to our reference experiment. If we take any final state vector
at time $t_{1}$ and propagate it back to $t_{0}$, we generate a
vector that must be a part of the full Hilbert space, but is not a
part of the physical subspace. The spin points in the wrong direction,
and the particle's wavefunction is not confined to the region indicated
by the first detector. Within the physical subspace, the particle
position is limited to the detected region. Further, as consequence
of the experimental set-up, if we continue to propagate back in time
within the physical subspace, the particle will have to have originated
at the location of the particle source. This means that, when restricted
to the physical subspace, the particle can only reach the first detector
by traveling through the SG set-up. Under these conditions, when the
first detector flashes, the particle spin must be in the y-direction.
This too must be reflected in the physical subspace. (This will be
ensured by a self-consistency condition that will be introduced in
Sec \ref{sec:The-condition}). Therefore, if the final state vector
is propagated back from $t_{1}$ to $t_{0}$ within the physical subspace,
the particle spin will point in the y-direction, and the particle
position will be consistent with the detector state. If the reference
experiment were to actually be performed, this is the state that we
know the experiment must have been in, given the final state. 

The physical subspace therefore appears to provide what is required
to resolve the issue described in the introduction. We can now turn
to drawing out some of its properties, and then utilizing it in the
Born rule.

(Side Note: The physical subspace will contain vectors in which the
particle spin points in the x-direction at time $t_{0}$, rather than
the y-direction. For example, those for SG devices that measure the
spin in the x-direction. But all of these are orthogonal to the given
final state of the measuring device at a time $t_{1}$, when propagated
back to $t_{0}$.)

\subsection{\label{subsec:The-physical-subspace}The physical subspace under
propagation \& time reversal}

Before continuing, we must briefly pause to take into account a ubiquitous
aspect of recordings. Something so omnipresent, we rarely make note
of it. It is this: we only ever have records of the past, we never
have a record of events that will occur in the future. We have abundant
information about events in one light cone, while the other is dark.

We ourselves remember the past, but have no memory of the future.
We can imagine having direct information about both what has happened
and what will happen, but in our world one of these types of knowledge
is forbidden to us. This does not only apply to our minds. When we
dig into the earth, we find evidence of past civilizations, and past
life forms, but find no evidence of future civilizations, or future
life forms. The effect can be seen in its most reductive form when
we peer into the night sky. As we look at distant stars, we see them
as they were in past, but we can not see how they will be in the future.

All of these can be attributed to the fact that we encounter retarded
waves, but do not encounter advanced waves. Retarded waves allow information
to bubble up from the past, but bring no information about the future.
Advanced waves would reverse this. If you wish, you can imagine that
unlike electromagnetism, gravitation does generate advanced waves,
and so when listening to gravitational waves we can detect (for example)
a collision of black holes that will happen in the future.

Here we will make no attempt to explain this asymmetry. We will simply
note its effect on the matters we wish to understand. 

Let's call a state ``regular'' if it does not contain any recordings
of future event (perhaps equivalently, a regular state does not contain
advanced waves), and limit $\mathcal{P}$ to the physical subspace
of regular states. A state, $\left|\psi\right\rangle $, is ''physically
possible'' if $\mathcal{P}\left|\psi\right\rangle =\left|\psi\right\rangle $.
If $\left|\psi\right\rangle $ is a regular, physically possible state,
the state we arrive at by propagating it forward in time should also
be a physically possible state. Thus, elements of $\mathcal{P}$ map
onto elements of $\mathcal{P}$ when propagated forward in time. For
example, in our reference experiment, if we take a physically possible
state at time $t_{0}$ (when the particle is detected after leaving
the first SG set-up), and propagate it forward to time $t_{1}$, we
get a linear combination of experimental outcome states. As experimental
outcome states must be physically possible, their linear combinations
are elements of the physical subspace.

However, if we propagate $\left|\psi\right\rangle $ backward in time,
the resulting vector might not be a physically possible state. Our
reference experiment gives an example of this. When a physically possible
final state at time $t_{1}$ is propagated back to time $t_{0}$,
the resulting state is not physically possible, because the spin points
in the wrong direction and the particle wavefunction is not confined
to the region of the first detector. These two aspects of the physical
subspace mean that, if the physical subspace is restricted to regular
states, then when $\mathcal{P}$ is propagated forward in time, it
will map onto a subspace of $\mathcal{P}$. Call this subspace $\mathcal{P}[\triangle t]$.
If $\mathcal{P}[\triangle t]$ is propagated back by $\triangle t$,
it will map back to $\mathcal{P}$. If $\mathcal{P}$ is propagated
back in time, it will map onto a superspace of $\mathcal{P}$.

These properties may seem odd, given that these subspaces are generated
by unitary transformations. They are allowed if both $\mathcal{P}$
and $\mathcal{P}^{\bot}$ are infinite, $\mathcal{P}^{\bot}$ being
the subspace orthogonal to $\mathcal{P}$. 

{[}This is fairly easy to prove. If $X$ and $Y$ are subspaces of
some Hilbert space, there exist unitary transformations from $X$
to $Y$ if and only if $X$ \& $Y$ have the same dimensionality,
and $X^{\bot}$ \& $Y^{\bot}$ also have the same dimensionality.
Clearly a unitary transformation can not exist if these dimensionality
constraints are not satisfied. To see that unitary transformations
must exist if they are satisfied, assume that these pairs have the
same dimensionality. Choose any orthonormal basis of $X$, $\left|x_{i}\right\rangle $,
and of $Y$, $\left|y_{i}\right\rangle $. Because these bases have
the same cardinality, there is a bijection between them. This also
holds for any orthonormal bases of $X^{\bot}$ \& $Y^{\bot}$, $\left|x_{i}^{\bot}\right\rangle $
\& $\left|y_{i}^{\bot}\right\rangle $. Take the combined bijection
from the combined basis $\left|x_{i}\right\rangle $ + $\left|x_{i}^{\bot}\right\rangle $
onto the combined basis $\left|y_{i}\right\rangle $ + $\left|y_{i}^{\bot}\right\rangle $.
Extend this mapping to the full Hilbert space by making it linear.
This yields a unitary mapping from $X$ onto $Y$.

If $X$ and $X^{\bot}$ are both infinite, and $Y$ is a subspace
of $X$ with the same cardinality as $X$, then $Y^{\bot}$ will be
a superspace of $X^{\bot}$ with the same cardinality as $X^{\bot}$.
As shown, under these conditions, unitary transformations from $X$
to $Y$ exist.{]} 

When composed entirely of regular states, $\mathcal{P}$ is not invariant
under time reversal. This is expected. Any set of states that only
contains retarded waves will not be invariant under time reversal.

Here, we will focus on regular states, as they correspond to our direct
experience. The physical subspace will be understood to be composed
of regular states.

From here on out, the Heisenberg picture will generally be employed.
In the Heisenberg picture, the dynamic properties of the physical
subspace can be summarized as: For any $t_{1}<t_{2}$, $\mathcal{P}(t_{1})\mathcal{P}(t_{2})=\mathcal{P}(t_{1})$. 

For any set of states, $X$, s.t. $\left[\mathbb{P}_{X}(t),\mathcal{P}(t)\right]=0$,
define $\mathcal{P}_{X}(t)=\mathcal{P}(t)\mathbb{P}_{X}(t)$. $\mathcal{P}_{X}(t)$
is the projection operator onto the portion of the physical subspace
for which $X$ holds true.

\subsection{\label{subsec:Moving-back-in}Moving back in time within the physical
subspace}

A set of states, $X$, is physically possible if $\left[\mathbb{P}_{X}(t),\mathcal{P}(t)\right]=0$
\& $\mathcal{P}(t)\mathbb{P}_{X}(t)\neq0$. If $X$ at $t_{f}$ is
physically possible, and we wish to accurately move $\mathcal{P}_{X}(t_{f})$
back in time, it may not be enough to simply propagate it back using
the propagator; we also need to trim away elements that are not part
of the physical subspace. For our reference experiment, this is how
we can move our outcome wavefunctions back in time and end up with
the particle's spin pointing in the proper direction as it exits the
first SG device.

Non-physical aspects of the wavefunction can be trimmed away by applying
$\mathcal{P}(t)$ at all prior times. However, due to the just described
properties of $\mathcal{P}$, to trim it at all times it is sufficient
to trim at just one time. For example if we were to trim $\mathcal{P}_{X}(t_{f})$
at times $t<t_{1}<....t_{n}<t_{f}$, we'd have $\mathcal{P}(t)\mathcal{P}(t_{1})...\mathcal{P}(t_{n})\mathcal{P}(t_{f})\mathbb{P}_{X}(t_{f})\mathcal{P}(t_{f})\mathcal{P}(t_{n})...\mathcal{P}(t_{1})\mathcal{P}(t)$.
By the properties of $\mathcal{P}$, that operator is equal to $\mathcal{P}(t)\mathbb{P}_{X}(t_{f})\mathcal{P}(t)$.
Thus, $\mathcal{P}(t)\mathbb{P}_{X}(t_{f})\mathcal{P}(t)$ alone trims
away all non-physical elements of $\mathbb{P}_{X}(t_{f})$ as we move
back in time from $t_{f}$ to $t$.

Quite commonly $\mathcal{P}(t)\mathbb{P}_{X}(t_{f})\mathcal{P}(t)$
will not be a projection operator; it is more akin to a density matrix.
It will be helpful to define the projection operator onto the smallest
subspace that contains $\mathcal{P}(t)\mathbb{P}_{X}(t_{f})\mathcal{P}(t)$.
Call this projection operator $\mathcal{P}_{(X,t_{f})}(t)$. It defined
so that, for any complete basis that diagonalizes $\mathcal{P}(t)\mathcal{\mathbb{P}}_{X}(t_{f})\mathcal{P}(t)$,
for any $\left|\psi;t\right\rangle $ in that basis, $\mathcal{P}_{(X,t_{f})}(t)\left|\psi;t\right\rangle =\left|\psi;t\right\rangle $
if $\mathcal{P}(t)\mathcal{P}_{X}(t_{f})\mathcal{P}(t)\left|\psi;t\right\rangle \neq0$,
and $\mathcal{P}_{(X,t_{f})}(t)\left|\psi;t\right\rangle =0$ otherwise.

\subsection{System1 \& System2}

To be able to readily describe the physical subspace, it will often
be helpful to partition closed systems into 2 subsystems, which we'll
call system1 \& system2. System1 will be the portion of the closed
system about whose state something will be asserted. In the physical
subspace, this assertion can imply something about the state of system2.
For example, system1 may include a detector, while system2 includes
the particles that can be detected; asserting something about the
state of detector may tell us something about the particle state.
Similarly, system1 may be you, and system2 may be the rest of the
universe. Asserting something about the state of your brain asserts
something about the state of the outside world.

One may imagine system1 \& system2 as being spatially divided, but
the don't need to be. We can more generally assume that for some complete
set of commuting observables for a closed system, $\vec{O}$, we divide
$\vec{O}$ into two subsets, $\vec{O}_{1}$ \& $\vec{O}_{2}$. System1
is the subspace corresponding to $\vec{O}_{1}$, and system2 is the
subspace of $\vec{O}_{2}$.

Unless stated otherwise, projection operators will project onto subspaces
of the full, closed system. If $X$ is any physically possible set
of system1 states, $\mathbb{P}_{X}(t)$ projects onto the Cartesian
product of the $X$ subspace in system1 and all of system2. $\mathcal{P}_{X}(t)$
restricts system2 states to those states that are physically consistent
with $X$.

\subsection{$X(t)$ }

If $X$ is a physically possible set of system1 observable values,
$\mathcal{P}_{X}(t_{f})\neq\mathcal{P}(t)\mathcal{P}_{X}(t_{f})\mathcal{P}(t)$
($t<t_{f}$) when $X$ records non-commuting measurements performed
on system2. For example, if $X$ contains information about a particle's
position at time $t_{1}$, and momentum at $t_{2}$. Or, as in our
reference example, a particle's spin along different directions. If
all measurements recorded in $X$ do commute, then $\mathcal{P}(t)\mathcal{P}_{X}(t_{f})\mathcal{P}(t)$
simply projects $\mathcal{P}_{X}(t_{f})$ onto vectors at time $t$.

However such cases raise another issue. If we wish to accurately portray
the information available at $t<t_{f}$ this may not be $\mathcal{P}(t)\mathcal{P}_{X}(t_{f})\mathcal{P}(t)$,
because this operator can reflect information gathered after $t$.
For example, if the energy state of an atom is measured at time $t_{0}$,
its angular momentum is measured at time $t_{\ensuremath{1}}$, and
the final state of the experimental equipment, $X$, implies both
values at time $t_{f}$, then within $\mathcal{P}_{X}(t_{f})$ the
atom will have the measured values for both energy \& angular momentum.
If we move $\mathcal{P}_{X}(t_{f})$ back in time to $t_{0}$ and
trim away non-physical aspects of the state, $\mathcal{P}(t_{0})\mathcal{P}_{X}(t_{f})\mathcal{P}(t_{0})$,
the atom will still have those values for energy and angular momentum.

To represent the information available at any given time, we will
need to derive from $\left(X,t_{f}\right)$ an $X(t)$ representing
the implied state of system1 at time $t<t_{f}$. $X$ is a range of
values for some set of system1 observables, $\vec{O}_{1}$. At each
$t$, $X(t)$ will also be a range of values of $\vec{O}_{1}$. These
$X(t)$ are generated as follows. Take the partial trace of $\mathcal{P}(t)\mathbb{P}_{X}(t_{f})\mathcal{P}(t)$
over system2, $Tr_{2}\left(\mathcal{P}(t)\mathbb{P}_{X}(t_{f})\mathcal{P}(t)\right)$.
Call this operator $\mathbb{A}_{1}$. $X(t)$ is the set of eigenvalues
of $\vec{O}_{1}$, $\vec{x}$, s.t. $\left\langle \vec{x}\right|\mathbb{A}_{1}\left|\vec{x}\right\rangle \neq0$.
$X(t)$ is then what is known with certainty about the system1's state
in the $\vec{O}_{1}$ basis at time $t$ ($t<t_{f}$), given $(X,t_{f})$.
If, as considered above, $X$ is the state of a recording, $X(t)$
is what is known about the state of the recording at an earlier time,
perhaps when only a portion of the events recorded in $X$ had been
recorded.

This mirrors our informal use of recordings. Using our own memories
as an example, if we remember observing something in the past, we
use that memory to project back to our mental state at the time the
observation was made, which we then use to deduce something about
the state of the outside world at the time of the observation. This
reasoning is generally reliable, and so we ought to be able to reproduce
it within quantum theory.

$X(t)$ will only be of value if $\left[\mathcal{P}(t),\mathbb{P}_{X(t)}(t)\right]=0$
for all $t$. If this does not hold, perhaps the wrong set of observables,
$\vec{O}_{1}$, was chosen. When $X(t)$ can be defined \& can be
used to describe past states, then it will be referred to as an ``observable
representation'' of past states.

\subsection{$T_{s}(X,t_{f})$}

For a given $\left(X,t_{f}\right)$, it will be useful to find the
time at which $X$ started gathering information.

First, consider times, $t$, that have the following property:

1) For all $t^{\prime}<t$: $\mathcal{P}(t^{\prime})\mathcal{P}_{X}(t_{f})\mathcal{P}(t^{\prime})=\mathcal{P}(t)\mathcal{P}_{X}(t_{f})\mathcal{P}(t)$ 

These are times at which $\left(X,t_{f}\right)$ had only accumulated
commuting information about the outside world. If there is an accepted
observable representation of past events, $X(t)$, and we wish to
find times prior to which $X$ also implies that commuting information
had not been gathered, we can add the further condition:

2) $\mathcal{P}_{X(t)}(t)=\mathbb{P}_{X(t)}(t)$

The times, $t$, that satisfy these two demands are times prior to
when $X$ started accumulating information about the outside world.
Take $T_{s}(X,t_{f})$ to be the least upper bound of the set of times
that satisfy (1) \& (2) (or just (1), if a satisfactory observable
representation can not be found). This is the time at which $X$ starts
accumulating information about the outside world. The ``s'' subscript
standing for ``start''. Property (1) will hold at time $T_{s}(X,t_{f})$,
though (2) may not.

As subspace $X$ gets larger, $T_{s}(X,t_{f})$ will get earlier.
If $X$ is chosen to be large enough, it is even possible that $T_{s}(X,t_{f})=-\infty$.
We will restrict ourselves to $X$'s for which $T_{s}(X,t_{f})$ is
finite.

We will soon be interested in operators of the form $\mathbb{P}_{X}(t)\mathcal{P}(t_{0})\mathbb{P}_{X}(t)$
($t>t_{0}$). For these, the following property will prove useful: 

For $t_{1}<t$ \& $t_{2}<t$, if $\mathcal{P}(t_{1})\mathbb{P}_{X}(t)\mathcal{P}(t_{1})=\mathcal{P}(t_{2})\mathbb{P}_{X}(t)\mathcal{P}(t_{2})$
then $\mathbb{P}_{X}(t)\mathcal{P}(t_{1})\mathbb{P}_{X}(t)=\mathbb{P}_{X}(t)\mathcal{P}(t_{2})\mathbb{P}_{X}(t)$.
From this it follows that, for all $t_{0}\leq T_{s}(X,t)$: $\mathbb{P}_{X}(t)\mathcal{P}(t_{0})\mathbb{P}_{X}(t)=\mathbb{P}_{X}(t)\mathcal{P}(T_{s}(X,t_{f}))\mathbb{P}_{X}(t)$. 

{[}Proof: Define $A(t^{\prime})=\mathbb{P}_{X}(t)\mathcal{P}(t^{\prime})\mathbb{P}_{X}(t)$
and $B(t^{\prime})=\mathcal{P}(t^{\prime})\mathbb{P}_{X}(t)\mathcal{P}(t^{\prime})$,
$A(t^{\prime})$ is Hermitian and non-negative. Therefore, $A(t_{1})=A(t_{2})$
iff $\left[A(t_{1})\right]^{2}=\left[A(t_{2})\right]^{2}$. $\left[A(t^{\prime})\right]^{2}=\mathbb{P}_{X}(t)\mathcal{P}(t^{\prime})\mathbb{P}_{X}(t)\mathcal{P}(t^{\prime})\mathbb{P}_{X}(t)=\mathbb{P}_{X}(t)B(t^{\prime})\mathbb{P}_{X}(t)$.
So if $B(t_{1})=B(t_{2})$ then $A(t_{1})=A(t_{2})$.{]}

\section{Probabilities within the physical subspace}

\subsection{\label{sec:The-condition}The condition}

We desire a form of the Born rule that eliminates problems like the
one that arises in our reference experiment. To arrive at it, we start
by considering the the form of the condition, or given quantity, as
it will appear in the the Born rule. 

Take the condition to be: At time $t_{c}$, observable statement $X$
holds true. A preliminary requirement is that $\left(X,t_{c}\right)$
must be physically possible, meaning $\left[\mathbb{P}_{X}(t_{c}),\mathcal{P}(t_{c})\right]=0$
\& $\mathcal{P}(t_{c})\mathbb{P}_{X}(t_{c})\neq0$. Take $t_{0}$
to be any $t_{0}\leq T_{s}(X,t)$. The condition corresponds to the
portions of vectors in physical subspace at $t_{0}$ that reach $X$
at $t_{c}$. The condition used in the Born rule is then: $\mathbb{P}_{X}(t_{c})\mathcal{P}(t_{0})\mathbb{P}_{X}(t_{c})$,
where $t_{0}\leq T_{s}(X,t)$. Because $\mathcal{P}(t_{0})=\mathcal{P}(t_{c})\mathcal{P}(t_{0})$,
this can also be written $\mathcal{P}_{X}(t_{c})\mathcal{P}(t_{0})\mathcal{P}_{X}(t_{c})$. 

If $X$ doesn't imply anything about observable values outside those
specified in $X$ (e.g., if $X$ does not imply any information about
system2), then $\mathcal{P}_{X}(t_{c})=\mathbb{P}_{X}(t_{c})$ and
$T_{s}(X,t_{c})=t_{c}$, so $\mathbb{P}_{X}(t_{c})\mathcal{P}(t_{0})\mathbb{P}_{X}(t_{c})$
is equal to $\mathbb{P}_{X}(t_{c})$. This is how the condition is
commonly represented.

$\mathbb{P}_{X}(t_{c})\mathcal{P}(t_{0})\mathbb{P}_{X}(t_{c})$ can
be expanded into the sequence events that are implied by $\left(X,t_{c}\right)$.
For simplicity, let's do this using an observable representation.
The $X(t)$ construction implies that for $t<t_{c}$, $\mathcal{P}_{X}(t_{c})$
is a subspace of $\mathbb{P}_{X(t)}(t)$, and so $\mathbb{P}_{X(t)}(t)\mathcal{P}_{X}(t_{c})=\mathcal{P}_{X}(t_{c})$.
Thus, for any $t_{0}<t_{1}<....t_{n}<t_{c}$: $\mathbb{P}_{X}(t_{c})\mathcal{P}(t_{0})\mathbb{P}_{X}(t_{c})=\mathcal{P}_{X}(t_{c})\mathcal{P}(t_{0})\mathcal{P}_{X}(t_{c})=\mathcal{P}_{X}(t_{c})\mathbb{P}_{X(t_{n})}(t_{n})...\mathbb{P}_{X(t_{0})}(t_{0})\mathcal{P}(t_{0})\mathbb{P}_{X(t_{0})}(t_{0})...\mathbb{P}_{X(t_{n})}(t_{n})\mathcal{P}_{X}(t_{c})$.
Because for all $t^{\prime}<t$, $\mathcal{P}(t^{\prime})\mathcal{P}(t)=\mathcal{P}(t^{\prime})$,
and because for all $t_{i}$, $\left[\mathbb{P}_{X(t_{i})}(t_{i}),\mathcal{P}(t_{i})\right]=0$,
this expression is $\mathbb{P}_{X}(t_{c})\mathcal{P}(t_{0})\mathbb{P}_{X}(t_{c})=\mathcal{P}_{X}(t_{c})\mathcal{P}_{X(t_{n})}(t_{n})...\mathcal{P}_{X(t_{0})}(t_{0})...\mathcal{P}_{X(t_{n})}(t_{n})\mathcal{P}_{X}(t_{c})$.
Using the same properties, we can now flip the final projection operators
back to $\mathbb{P}_{X}(t_{c})$: $\mathbb{P}_{X}(t_{c})\mathcal{P}(t_{0})\mathbb{P}_{X}(t_{c})=\mathbb{P}_{X}(t_{c})\mathcal{P}_{X(t_{n})}(t_{n})...\mathcal{P}_{X(t_{0})}(t_{0})...\mathcal{P}_{X(t_{n})}(t_{n})\mathbb{P}_{X}(t_{c})$.

The times at which information had been gathered can be deduced based
on how $\mathcal{P}_{X(t)}(t)$ changes with time. Taking those to
be the times in $\mathbb{P}_{X}(t_{c})\mathcal{P}_{X(t_{n})}(t_{n})...\mathcal{P}_{X(t_{0})}(t_{0})...\mathcal{P}_{X(t_{n})}(t_{n})\mathbb{P}_{X}(t_{c})$,
we arrive at the usual form for the quantum condition, when conditioning
on the outcomes from a sequence of measurements. Here, the given information
can be viewed as a recording of a sequence of measurements that system1
performed on system2 at times $t_{0},t_{1},...,t_{n}$, with known
outcomes captured by $X(t_{0})$, $X(t_{1})$, etc.

If $t_{n}$ is the time of the final measurement recorded in $X$,
and $t_{c}>t_{n}$, then the outer factors of $\mathbb{P}_{X}(t_{c})$
contain information about system1 that might be lacking in $\mathcal{P}_{X(t_{n})}(t_{n})$.
If we are only interested in system2 probabilities, then the $\mathbb{P}_{X}(t)$
will play no role. Very commonly, we would condition on the state
at $t_{n}$, in which case that projection operator can be eliminated
from the expression.

{[}Side Note: There is no necessity to use an observable representation.
We can instead expand the condition operator in terms of $\mathcal{P}_{(X,t_{c})}(t)$
rather than $\mathcal{P}_{X(t)}(t)$. (If the usage ``$\mathcal{P}_{(X,t_{c})}(t)$''
doesn't ring a bell, you can refer back to Sec \ref{subsec:Moving-back-in}
for it's definition.) This follows from the fact that for any $\left|\psi;t_{c}\right\rangle $
in the $\mathbb{P}_{X}(t_{c})$ subspace, $t<t_{c}$, $\mathcal{P}_{(X,t_{c})}(t)\left(\mathcal{P}(t)\left|\psi;t_{c}\right\rangle \right)=\mathcal{P}(t)\left|\psi;t_{c}\right\rangle $.
Because $\mathbb{P}_{X}(t_{c})\mathcal{P}(t_{0})\mathbb{P}_{X}(t_{c})=\mathbb{P}_{X}(t_{c})\mathcal{P}(t)\mathcal{P}(t_{0})\mathcal{P}(t)\mathbb{P}_{X}(t_{c})$
($t_{c}>t>t_{0}$), this yields $\mathbb{P}_{X}(t_{c})\mathcal{P}(t_{0})\mathbb{P}_{X}(t_{c})=\mathbb{P}_{X}(t_{c})\mathcal{P}_{(X,t_{c})}(t)\mathcal{P}(t_{0})\mathcal{P}_{(X,t_{c})}(t)\mathbb{P}_{X}(t_{c})$.
Repeating this operation, we get, for any $t_{0}<t_{1}<....t_{n}<t_{c}$,
$\mathbb{P}_{X}(t_{c})\mathcal{P}(t_{0})\mathbb{P}_{X}(t_{c})=\mathbb{P}_{X}(t_{c})\mathcal{P}_{(X,t_{c})}(t_{n})...\mathcal{P}_{(X,t_{c})}(t_{0})...\mathcal{P}_{(X,t_{c})}(t_{n})\mathbb{P}_{X}(t_{c})$.{]}

The condition $\left(X,t_{c}\right)$ can be represented by $\mathbb{P}_{X}(t)\mathcal{P}(t_{0})\mathbb{P}_{X}(t)$
under all circumstances, not just the textbook example of a sequence
of instantaneous measurements. For example, $\mathbb{P}_{X}(t)\mathcal{P}(t_{0})\mathbb{P}_{X}(t)$
will hold when $X$ contains information that was continuously gained,
rather than only being gained at discrete times. It will also hold
on cases where it is known that something must have happened, but
it is unclear when it happened. Take, for example, our reference SG
experiment, but remove the first detector. If $X$ is a recording
of the experimental run through the outcome, then given $X$ it is
clear that the particle evaded the barrier, at which point its spin
was in the y-direction, but it would be unclear when it passed by
barrier. If $t_{0}$ is the start of experiment, then starting at
$\mathcal{P}(t_{0})$ and conditioning on the fact that the set-up
must be in $X$ at time $t$, the parts of the wavefunctions that
can not reach $X$ at $t$ are continuously trimmed away as we move
from $t_{0}$ to $t$. This trims away any bit of any wavefunction
that hits the barrier. $\mathbb{P}_{X}(t)\mathcal{P}(t_{0})\mathbb{P}_{X}(t)$
therefore ensures that at $\left(X,t\right)$ it is known that the
particle evaded the barrier, at which point its spin was in the y-direction,
even if it is not known when this happened. 

\subsubsection*{A note on the make up of the physical subspace }

In Sec \ref{subsec:Introducing-the-physical} it was mentioned that,
given the states of the experimental equipment in our reference experiment,
in the physical subspace the particle's spin must point in the y-direction
at $t_{0}$, and in the x-direction at $t_{1}$. This conclusion is
borne out by experience. To arrive at it, the physical subspace must
take into account information that has been retained about past events.
This allows the physical subspace to reflect both that the particle
is in the location of where a detector flashed, and that it got there
by originating at the particle source and moving through the SG device(s).
Using this information, the detected position also determines the
particle's spin.

Because the physical subspace takes into account information about
past events that are implied by the current state, and because this
information is encapsulated in $\mathcal{P}_{X}(t)\mathcal{P}(t_{0})\mathcal{P}_{X}(t)$,
we can expect the physical subspace to obey the following self-consistency
condition: If $\left[\mathbb{P}_{X}(t),\mathcal{P}(t)\right]=0$,
$\mathcal{P}_{X}(t)$ will be the projection operator onto $\mathcal{P}_{X}(t)\mathcal{P}(t_{0})\mathcal{P}_{X}(t)$'s
subspace. $\mathcal{P}_{X}(t)\mathcal{P}(t_{0})\mathcal{P}_{X}(t)$
adds information about the probabilities of the various states within
$\mathcal{P}_{X}(t)$, but they both span the same states.

\subsection{\label{sec:The-Born-rule}The Born rule}

The Born rule follows more or less immediately from the condition.
One open question is how to choose $t_{0}$. Let's call the complete
set of outcomes who's probabilities we wish to calculate $\varUpsilon$
(e.g., $\varUpsilon$ is the set of possible outcomes of a measurement).
If the elements of $\varUpsilon$ contain information about past events,
and for each $Y\in\varUpsilon$ we wish to be able to expand $\mathbb{P}_{Y}(t)$
into its sequence of implied events, we take $t_{0}$ to be any lower
bound of $\left\{ T_{s}(X,t_{c})\right\} \bigcup\left\{ T_{s}(Y,t):Y\in\varUpsilon\right\} $,
where $\left(X,t_{c}\right)$ is the condition. In all other cases,
$t_{0}$ remains any time less than of equal to $T_{s}(X,t_{c})$.

With that, the amended Born rule is:

If $t\geq t_{c}$:

\[
P(Y,t|X,t_{c})=\frac{1}{Tr\left(\mathbb{P}_{X}(t_{c})\mathcal{P}(t_{0})\right)}Tr\left(\mathbb{P}_{Y}(t)\mathbb{P}_{X}(t_{c})\mathcal{P}(t_{0})\mathbb{P}_{X}(t_{c})\right)
\]

If $t_{c}>t>t_{0}$, we would like to continue to be able to expand
$\mathbb{P}_{X}(t_{c})\mathcal{P}(t_{0})\mathbb{P}_{X}(t_{c})$ into
the time sequence of events that are implied by the condition, and
insert $\left(\varUpsilon,t\right)$ into that sequence. Ensuring
this in all generality requires the unpretty rule:

\[
P(Y;t|X;t_{c})=\frac{1}{\mathcal{N}}Tr\left(\mathbb{P}_{X}(t_{c})\mathcal{P}(t)\mathbb{P}_{Y}(t)\mathcal{P}_{(X,t_{c})}(t)\mathcal{P}(t_{0})\mathcal{P}_{(X,t_{c})}(t)\mathbb{P}_{Y}(t)\mathcal{P}(t)\right)
\]

where $\mathcal{N}=\sum_{Z\in\Upsilon}Tr\left(\mathbb{P}_{X}(t_{c})\mathcal{P}(t)\mathbb{P}_{Z}(t)\mathcal{P}_{(X,t_{c})}(t)\mathcal{P}(t_{0})\mathcal{P}_{(X,t_{c})}(t)\mathbb{P}_{Z}(t)\mathcal{P}(t)\right)$.

The $\mathcal{P}(t)$'s ensure that $X$'s implied sequence of events
are taken into account from $t$ to $t_{c}$; the $\mathcal{P}_{(X,t_{c})}$
ensures they are taken into account from $t_{0}$ to $t$. If $\left[\mathcal{P}(t),\mathbb{P}_{Y}(t)\right]=0$
then the $\mathcal{P}(t)$'s can be removed from the expression, as
they will be implied by the other factors. Similarly, if $\left[\mathcal{P}_{(X,t_{c})}(t),\mathbb{P}_{Y}(t)\right]=0$,
then the $\mathcal{P}_{(X,t_{c})}(t)$'s can be removed.

A far simpler form for $t_{c}>t>t_{0}$ arises if we are interested
in the probability of $\left(Y,t\right)$ based only on what was known
at $t$ given $\left(X,t_{c}\right)$. This is either:

\[
P(Y;t|X;t_{c})=\frac{1}{Tr\left(\mathcal{P}_{(X,t_{c})}(t)\mathcal{P}(t_{0})\right)}Tr\left(\mathbb{P}_{Y}(t)\mathcal{P}_{(X,t_{c})}(t)\mathcal{P}(t_{0})\mathcal{P}_{(X,t_{c})}(t)\right)
\]

or 

\[
P(Y;t|X;t_{c})=\frac{1}{Tr\left(\mathbb{P}_{X(t)}(t)\mathcal{P}(t_{0})\right)}Tr\left(\mathbb{P}_{Y}(t)\mathbb{P}_{X(t)}(t)\mathcal{P}(t_{0})\mathbb{P}_{X(t)}(t)\right)
\]

depending on the particular requirements. Note that in both the first
of these equations and the larger, ``unpretty'', expression for
$P(Y;t|X;t_{c})$, the space of states at time $t$, given $\left(X,t_{c}\right)$,
are the same. The larger expression may contain information about
the state probabilities that was gathered after $t$, which would
not be in the smaller expression.

Finally, there may be cases in which $\varUpsilon$, $t$, and $t_{0}$
are chosen so that $t\leq t_{0}$. In these cases:

\[
P(Y;t|X;t_{c})=\frac{1}{Tr\left(\mathbb{P}_{X}(t_{c})\mathcal{P}(t_{0})\right)}Tr\left(\mathbb{P}_{X}(t_{c})\mathcal{P}(t_{0})\mathbb{P}_{Y}(t)\mathcal{P}(t_{0})\right)
\]
.

This form may be used as an approximation when $t_{c}>t>t_{0}$, by
making $t_{0}=t$:

\[
P(Y;t|X;t_{c})=\frac{1}{Tr\left(\mathbb{P}_{X}(t_{c})\mathcal{P}(t)\right)}Tr\left(\mathcal{P}(t)\mathbb{P}_{X}(t_{c})\mathcal{P}(t)\mathbb{P}_{Y}(t)\right)
\]
. 

$\mathcal{P}(t)\mathbb{P}_{X}(t_{c})\mathcal{P}(t)$ is $\mathcal{P}_{X}(t_{c})$
propagated back in time within the physical subspace to $t$, and
$\frac{1}{\mathcal{N}}\mathcal{P}(t)\mathbb{P}_{X}(t_{c})\mathcal{P}(t)$
is the density matrix corresponding to $\mathbb{P}_{X}(t_{c})$ propagated
back to $t$. The three ways of forming a density matrix for the condition
at $t$ when $t<t_{c}$ are interrelated: $\mathcal{P}_{(X,t_{c})}(t)$
is derived from $\mathcal{P}(t)\mathbb{P}_{X}(t_{c})\mathcal{P}(t)$,
and given the basis for $X(t)$, $\mathbb{P}_{X(t)}(t)$ is derived
from $\mathcal{P}_{(X,t_{c})}(t)$.

It has implicitly been assumed that the Born rule, as stated here,
applies to closed systems. For example, when calculating the probability
of the outcome of an experiment, we do not only consider the system
being experimented on, but also the measuring devices, experimenters,
etc. The probabilities for an open system (e.g., just the system being
experimented on) can be derived from the probabilities for the larger
closed system. This will be further discussed in the next section.
When applied to closed systems, these forms of the Born rule are complete;
no other form is ever needed.

It should also be noted that the Born rule isn't only applicable to
actual observations or measurements. It is equally applicable to a
case in which we are interested in the probability spectrum of some
quantity, even if it is not measured. This distinction is of some
importance, because the physical subspace implies which observations
can be performed. The Born rule may be both applicable \& useful,
even in cases where the physical subspace dictates that under the
given conditions (described by $X$) observers that would allow the
observation to take place are not present.

\section{Measurement processes}

Let's now apply the Born rule to measurement processes. 

Before proceeding, we note that for a measurement process there are
two ways to split the closed system into system1 \& system2. Each
has its advantages.

In the first way, system1 contains experimental equipment (measuring
devices, recording devices, equipment for preparing the initial state,
etc.), while system2 contains the system to be measured. This is the
division that will be used here.

Alternately, system1 can contain only the recording device, including
its recording. System2 contains everything else. System2 can then
itself be subdivided into systemE, which includes system2's experimental
equipment, such as the measuring devices, and systemS, which contains
the system being experimented on (system2 of the prior division).
In the physical subspace, the state of system1 implies something about
the state of systemE, and the state of systemE implies something about
the state of systemS. That is, the recording records the measurement
preparation and outcome, which implies something about the state of
the system being measured. In this division, system1 essentially plays
the role of the external observer. In some ways this division is richer,
but it also introduces an extra layer of implication. The first division
will therefore be used.

\subsection{General measurements}

A measurement process has a space of system1 start states, $M_{0}$,
and a set of spaces for system1 end states, $M_{i}$ ($i=1,2,...$).
Within $M_{0}$ all measurement equipment are operational, and a record
is preserved of the preparation of system2. Call the measurement start
time $t_{1}$. Due to the recording of the preparation, $\mathbb{P}_{M_{0}}(t_{1})\mathcal{P}(t_{0})\mathbb{P}_{M_{0}}(t_{1})$
contains information about system2's initial state. Each system1 end
space, $M_{i}$, corresponds to a measurement outcome. In each of
these end spaces, the measuring equipment is in a state that reveals
the outcome of the measurement. The measurement completes by $t_{2}$,
and information about the outcome is preserved in system1, along with
the information about the preparation.

For any given outcome, $M_{i}$, and for each $t_{1}\leq t\leq t_{2}$,
we can construct an operator, $\hat{\kappa}_{i}(t)$, that encapsulates
was known about system2 at $t$. If we are using an observable representation,
$M(t)$, then 
\[
\hat{\kappa}_{i}(t)=\frac{1}{Tr(\mathbb{P}_{M_{0}}(t_{1})\mathcal{P}(t_{0}))}Tr_{1}(\mathbb{P}_{M_{i}(t)}(t)\mathbb{P}_{M_{0}}(t_{1})\mathcal{P}(t_{0})\mathbb{P}_{M_{0}}(t_{1})\mathbb{P}_{M_{i}(t)}(t))
\]
 Where ``$Tr_{1}$'' is the partial trace over system1. $M_{i}(t)$
is the space of observable states that the system1 must have been
in at $t$, given the the outcome is $M_{i}$. The partial trace means
that $\hat{\kappa}_{i}(t)$ represents all of the observable probabilities
of system2, given that system1 is in $M(t)$.

If an observable representation is not used then 
\[
\hat{\kappa}_{i}(t)=\frac{1}{Tr(\mathbb{P}_{M_{0}}(t_{1})\mathcal{P}(t_{0}))}Tr_{1}(\mathcal{P}_{(M_{i},t_{1})}(t)\mathbb{P}_{M_{0}}(t_{1})\mathcal{P}(t_{0})\mathbb{P}_{M_{0}}(t_{1})\mathcal{P}_{(\mathtt{M}_{i},t_{1})}(t))
\]

$\hat{\kappa}_{i}(t)$ is essentially the system2 density matrix at
time $t$ as the measurement unfolds and system1 gathers information
about sytem2. The only difference between $\hat{\kappa}_{i}(t)$ and
the density matrix, $\hat{\rho}_{i}(t)$, is the normalization: $\hat{\rho}_{i}(t)=\frac{1}{Tr(\hat{\kappa}_{i}(t))}\hat{\kappa}_{i}(t)$.
Unlike the density matrix, $\hat{\kappa}_{i}(t_{2})$ is normalized
to the probability that $M_{i}$ will be the outcome: $P(M_{i};t_{2}|M_{0};t_{1})=Tr(\hat{\kappa}_{i}(t_{2}))$.
Thus $\hat{\kappa}_{i}(t)$ gives us both what is known about the
state of system2 as the experiment unfolds, and the probability that
the outcome will be $M_{i}$.

The outcome, $M_{i}$, uniquely determines the system2 path, $\hat{\kappa}_{i}(t)$.
For measurement processes this goes both ways: the system2 density
matrix path, $\hat{\rho}_{i}(t)=\frac{1}{Tr(\hat{\kappa}_{i}(t))}\hat{\kappa}_{i}(t)$,
across $t_{1}\leq t\leq t_{2}$ is sufficient to determine the outcome,
$M_{i}$. That is, the outcome is determined entirely by the information
that system1 has about system2. This is the defining property of measurements.
For non-measurement processes, $M_{i}$ can also reflect information
about system1 that is not contained to system2.

It follows that a measurement process can be represented entirely
in terms of system2 using $\hat{\kappa}_{i}(t)$. The $\hat{\kappa}_{i}(t)$
paths are sufficient to determine the outcome, the outcome probability,
and what is known about system2 at all times. The particulars of the
measuring devices, the method of preparation, and so forth can be
set aside; all that matters is that they lead to the required set
of system2 $\hat{\kappa}_{i}(t)$ paths.

Starting at $t_{1}$ and moving forward to $t_{2}$, $\hat{\kappa}(t)$
evolves stochastically. If the final measurement outcome represents
a sequence of individual measurements performed at at discrete times,
$\hat{\kappa}(t)$'s time evolution is deterministic except for non-deterministic
jumps at those discrete times, reflecting the outcomes of the sequence
of measurements. For continuous measurements, $\hat{\kappa}(t)$ executes
a continuous random walk. As previously noted in Sec \ref{sec:The-condition},
when $M_{i}$ implies a measured value, but does not imply a particular
time at which the measurement took place, $\hat{\kappa}(t)$ will
be continuously trimmed to eliminate portions that are inconsistent
with the measured value. Finally, at times when $\hat{\rho}(t)$'s
quantum potential is small, its path will be consistent with classical
mechanics.

Thus, after making the minor amendments to the Born rule described
in the prior section, the theory can be used to present a unified
framework for all types of measurements. This would not have been
possible without those amendments, because the measurement description
requires accurately moving the possible outcomes back in time, and
deducing the system2 state based on the system1 state. Neither of
those operations are possible without the physical subspace. 
\begin{rem*}
As things stand, for a given measurement outcome $M_{i}$, any eigenvalues
$l,m\in M_{i}$, it may be the case that $\hat{\rho}_{m}(t)\neq\hat{\rho}_{l}(t)$
at some time $t$. We can always subdivide our $M_{i}$ into into
equivalence classes, $\mathtt{M}_{x}$, such that for a given $x$,
and any $l,m\in\mathtt{M}_{x}$, $\hat{\rho}_{m}(t)=\hat{\rho}_{l}(t)$
for all $t_{0}\leq t\leq t_{1}$. The probabilities for the outcomes,
$M_{i}$, can then be calculated by summing over the $\mathtt{M}_{x}$
that are within $M_{i}$. These $\mathtt{M}_{x}$ are the smallest
sets that can be considered measurement outcomes (if we subdivide
them further, the outcomes will not satisfy the definition of a measurement),
and measurement outcomes can be any partition of these $\mathtt{M}_{x}$.
\end{rem*}

\subsection{An example: tracking detectors}

Tracking detectors represent an interesting measurement process because
the measurement unfolds continuously, and because they reveal stable
particle paths, something that quantum theory may appear to rule out.

If $M_{i}$ is the final measurement state, containing a recording
of a particle passing through the tracking device, take $\hat{\rho}_{i}(t)$
to be the density matrix path for the particle derived from $M_{i}$
as described above. The position probability distribution for the
particle location as it passes through the tracking detector is then
$\varrho(\vec{x},t)=Tr\left(\mathbb{P}_{\vec{x}}(t)\hat{\rho}_{i}(t)\right)$.
$\varrho(\vec{x},t)$ should match our intuitive analysis of particle
position as a function of time, based on the final recording.

The change in $\varrho(\vec{x},t)$ with time goes a long way towards
determining the probability current. Under maximal information, including
known spin, the probability distribution as a function of time wholly
determines the probability current (e.g., if spin is constant in space,
and in the absence of a magnetic field, $\nabla\cdot\vec{j}=\frac{\partial}{\partial t}\rho$
\& $\nabla\times\vec{j}=\frac{1}{\rho}\nabla\rho\times\vec{j}$),
and these two determine the particle's wavefunction up to a meaningless
overall phase 

Tracking devices thus determine the particle's probability current,
rather than their instantaneous velocity or momentum. This has nothing
to do with the particulars of quantum theory; if particles obey classical
mechanics rather than quantum mechanics, tracking detectors would
still yield their position probability distributions as a function
of time, and with that, their probability currents (up to a curl).
Within quantum theory, this is significant because at a given time
any probability distribution can be paired with any probability current.
Tracking detectors therefore do not challenge the uncertainty principle.
They still, however, yield information about momentum: non-relativistically,
the spacial integral over the probability current, multiplied by the
particle's mass, is the momentum expectation value.

As noted above, if $\hat{\rho}_{i}(t)$'s quantum potential is small
compared to the interaction with the detector then we can expect the
$\varrho(\vec{x},t)$ to be consistent with classical paths. Thus,
all aspects of tracking detector's observed outcomes are a consequence
of the modified Born rule.

\section{Verifiable probabilities}

As the Born rule is an aspect of a scientific theory, it is fair to
ask: Under what circumstances can we test whether the predicted probabilities
are valid? There would seem to be three requirements. The first is
that the underlying dynamics enable reproducibility. Quantum mechanics
provides this. Reproducibility is a well known requirement, and will
not be further delved into here. The second is that, if $\varUpsilon$
is the outcome set, then all $Y_{i}\in\varUpsilon$, $\left[\mathcal{\mathbb{P}}_{Y_{i}}(t),\mathcal{P}(t)\right]=0$.
Without this, the outcomes couldn't be obtained. The third is that
all relevant information must be retained. This is required in order
for the statistical probability to be calculated. When these requirements
are satisfied, the probabilities predicted by the the Born rule are
``verifiable''. The remainder of this section will investigate the
last requirement. Its nature will be slightly different for $t>t_{c}$
\& $t_{c}>t$ ($t_{c}$ being the time of the condition).

\subsection{For $t>t_{c}$}

When $t>t_{c}$, we require that when any outcome, $Y_{i}\in\varUpsilon$,
is obtained at time $t$, it retains the information that the condition,
$X$, occurred at $t_{c}$. Without this, there would be no way to
gather data on how likely the outcomes $Y_{i}\in\varUpsilon$ are,
given $X$. For experiments, this is also a necessary condition for
reproducibility; an experiment can not be reproduced if its set-up
is uncertain, and if this requirement is not upheld then at the end
of the experiment, $\left(Y_{i},t\right)$, the set-up, $\left(X,t_{c}\right)$,
is no longer recalled with certainty.

Let's state this requirement more precisely. As discussed in Sec \ref{subsec:Moving-back-in},
given $\left(Y_{i},t\right)$, $\mathcal{P}(t_{c})\mathcal{\mathbb{P}}_{Y_{i}}(t)\mathcal{P}(t_{c})$
is what is known about the closed system at time $t_{c}$. The demand
that the outcomes retain information about the condition therefore
requires $\mathcal{P}(t_{c})\mathcal{\mathbb{P}}_{Y_{i}}(t)\mathcal{P}(t_{c})$
to commute with $\mathcal{P}_{X}(t_{c})$. This can be written $\mathcal{P}(t_{c})\left[\mathcal{\mathbb{P}}_{Y_{i}}(t),\mathcal{\mathbb{P}}_{X}(t_{c})\right]\mathcal{P}(t_{c})=0$,
or equivalently, $\mathcal{P}(t_{c})\left[\mathcal{P}_{Y_{i}}(t),\mathcal{P}_{X}(t_{c})\right]\mathcal{P}(t_{c})=0$.
To draw out what this requirement entails, a couple of projection
operators to go along with $\mathcal{P}_{Y_{i}}(t)$ will prove helpful. 

\subsection{$\mathcal{P}_{Z_{i}}(t)$ \& $\mathcal{P}_{W_{i}}(t)$}

For each $Y_{i}\in\varUpsilon$, let's define $\mathcal{P}_{Z_{i}}(t)$
to be the projection operator onto the elements of the $\mathcal{P}_{Y_{i}}(t)$
subspace that must have come from $\mathcal{P}_{X}(t_{c})$. Propagating
$\left|\phi,t\right\rangle $ back to $t_{c}$ within the physical
subspace yields $\mathcal{P}(t_{c})\left|\phi,t\right\rangle $ (we
can ignore normalization). This mean that $\mathcal{P}_{Z_{i}}(t)$
spans the space of states, $\left|\phi,t\right\rangle $, s.t. $\mathcal{P}_{Y_{i}}(t)\left|\phi,t\right\rangle =\left|\phi,t\right\rangle $,
$\mathcal{P}(t_{c})\left|\phi,t\right\rangle \neq0$, and $\mathcal{P}_{X}(t_{c})\left(\mathcal{P}(t_{c})\left|\phi,t\right\rangle \right)=\left(\mathcal{P}(t_{c})\left|\phi,t\right\rangle \right)$. 

It may be interesting to note that we can know that $\left|\phi,t\right\rangle $
must have come from $\mathcal{P}_{X}(t_{c})$ without it being in
the $\mathcal{P}_{X}(t_{c})$ subspace. This is because only $\mathcal{P}(t_{c})\left|\phi,t\right\rangle $
needs to be in $\mathcal{P}_{X}(t_{c})$, $\left(1-\mathcal{P}(t_{c})\right)\left|\phi,t\right\rangle $
does not.

Defining $\mathcal{\mathbb{P}}_{\neg X}(t_{c})=1-\mathcal{\mathbb{P}}_{X}(t_{c})$,
we can similarly define $\mathcal{P}_{W_{i}}(t)$ to be the projection
operator onto the space of states s.t. $\mathcal{P}_{Y_{i}}(t)\left|\phi,t\right\rangle =\left|\phi,t\right\rangle $
and $\mathcal{P}_{\neg X}(t_{c})\left(\mathcal{P}(t_{c})\left|\phi,t\right\rangle \right)=\left(\mathcal{P}(t_{c})\left|\phi,t\right\rangle \right)$.
$\mathcal{P}_{W_{i}}(t)$ consists of the elements of $\mathcal{P}_{Y_{i}}(t)$
that we know for sure did not come from $\mathcal{P}_{X}(t_{c})$.
$\mathcal{P}_{Z_{i}}(t)$ \& $\mathcal{P}_{W_{i}}(t)$ are disjoint
subspaces, but they need not be orthogonal; only $\mathcal{P}_{(Z_{i},t)}(t_{c})$
\& $\mathcal{P}_{(W_{i},t)}(t_{c})$ must be orthogonal.

When $\mathcal{P}(t_{c})\left[\mathcal{\mathbb{P}}_{Y_{i}}(t),\mathcal{\mathbb{P}}_{X}(t_{c})\right]\mathcal{P}(t_{c})=0$,
there is a complete set of eigenstates of $\mathcal{P}(t_{c})\mathcal{P}_{Y_{i}}(t)\mathcal{P}(t_{c})$
that can be divided into those entirely within $\mathcal{P}_{X}(t_{c})$,
and those orthogonal to $\mathcal{P}_{X}(t_{c})$. Under these conditions,
$\mathcal{P}(t_{c})\mathcal{P}_{Z_{i}}(t)\mathcal{P}(t_{c})=\mathcal{P}_{X}(t_{c})\left[\mathcal{P}(t_{c})\mathcal{P}_{Y_{i}}(t)\mathcal{P}(t_{c})\right]$
and $\mathcal{P}(t_{c})\mathcal{P}_{W_{i}}(t)\mathcal{P}(t_{c})=\mathcal{P}_{\neg X}(t_{c})\left[\mathcal{P}(t_{c})\mathcal{P}_{Y_{i}}(t)\mathcal{P}(t_{c})\right]$.
Together, these mean that $\mathcal{P}(t_{c})\mathcal{P}_{Y_{i}}(t)\mathcal{P}(t_{c})=\mathcal{P}(t_{c})\mathcal{\mathbb{P}}_{Z_{i}}(t)\mathcal{P}(t_{c})+\mathcal{P}(t_{c})\mathcal{\mathbb{P}}_{W_{i}}(t)\mathcal{P}(t_{c})$.
By the definition of $W_{i}$, $\mathcal{\mathbb{P}}_{X}(t_{c})\left[\mathcal{P}(t_{c})\mathcal{\mathbb{P}}_{W_{i}}(t)\mathcal{P}(t_{c})\right]=0$,
so when the commutation relation holds, $Tr(\mathcal{\mathbb{P}}_{Y_{i}}(t)\mathcal{\mathbb{P}}_{X}(t_{c})\mathcal{P}(t_{0})\mathcal{\mathbb{P}}_{X}(t_{c}))=Tr(\mathcal{P}_{Z_{i}}(t)\mathcal{P}(t_{0}))$.

Putting this together: For $t>t_{c}$ the Born rule is verifiable
if, for all $Y_{i}\in\varUpsilon$: (1) $\left[\mathcal{\mathbb{P}}_{Y_{i}}(t),\mathcal{P}(t)\right]=0$
\& (2) $\mathcal{P}(t_{c})\left[\mathcal{\mathbb{P}}_{Y_{i}}(t),\mathcal{\mathbb{P}}_{X}(t_{c})\right]\mathcal{P}(t_{c})=0$.
When these hold, the Born rule can be written: $P(Y_{i};t|X;t_{c})=\frac{1}{Tr(\mathcal{P}_{X}(t_{c})\mathcal{P}(t_{0}))}Tr(\mathcal{P}_{Z_{i}}(t)\mathcal{P}(t_{0}))$.
Under these conditions, the only possible states of $\left(Y_{i},t\right)$
given $\left(X,t_{c}\right)$ are those for which it is known that
$X$ must have occurred at $t_{c}$. 

\subsection{For $t<t_{c}$}

For $t<t_{c}$ we want to know, following an outcome of $\varUpsilon$
being obtained at $t$, what information can be gathered at $t_{c}$
that will allow us to retain knowledge of the outcome.

It follows from our prior analysis that this requires, for all $Y_{i}$:
$\left[\mathcal{\mathbb{P}}_{Y_{i}}(t),\mathcal{P}(t)\right]=0$ \&
$\mathcal{P}(t)\left[\mathcal{\mathbb{P}}_{Y_{i}}(t),\mathcal{\mathbb{P}}_{X}(t_{c})\right]\mathcal{P}(t)=0$.
These are the conditions for verifiable probabilities when $t<t_{c}$.
$\frac{1}{Tr(\mathcal{P}_{X}(t_{c})\mathcal{P}(t))}\mathcal{P}(t)\mathcal{P}_{X}(t_{c})\mathcal{P}(t)$
is the density matrix for what is known at time $t$, given $X$ at
$t_{c}$, so the second condition, $\mathcal{P}(t)\left[\mathcal{\mathbb{P}}_{Y_{i}}(t),\mathcal{\mathbb{P}}_{X}(t_{c})\right]\mathcal{P}(t)=0$,
may be viewed as ensuring that the individual $Y_{i}$ are non-destructive
measurements on this density matrix.

For $t_{0}<t<t_{c}$ verifiable probabilities, the ``unpretty''
Born rule simplifies to \\
$P(Y_{i},t;X,t_{c})=\frac{1}{Tr(\mathbb{P}_{X}(t_{c})\mathcal{P}(t_{0}))}Tr\left(\mathcal{P}_{Y_{i}}(t)\mathbb{P}_{X}(t_{c})\mathcal{P}(t_{0})\right)$.

For each $Y_{i}$, $Z_{i}$ can be defined similarly to the prior
case; $Z_{i}$ consists of the elements of $X$ that must have come
from $Y_{i}$. When probabilities are verifiable, $\mathcal{P}(t)\mathcal{P}_{Z_{i}}(t_{c})\mathcal{P}(t)=\mathcal{P}_{Y_{i}}(t)\mathcal{P}_{X}(t_{c})\mathcal{P}(t)$,
so \\
$P(Y_{i},t;X,t_{c})=\frac{1}{Tr(\mathbb{P}_{X}(t_{c})\mathcal{P}(t_{0}))}Tr\left(\mathcal{P}_{Z_{i}}(t_{c})\mathcal{P}(t_{0})\right)$.
Same as the $t>t_{c}$ case.

Given $\left(X,t_{c}\right)$, some $\left(Y_{i},t\right)$ may not
be possible. For those $\left(Y_{i},t\right)$, there is no $Z_{i}$.
For all the other $\left(Y_{i},t\right)$, we can condition on both
$\left(X,t_{c}\right)$ and the $\left(Y_{i},t\right)$ outcome by
conditioning on $\left(Z_{i},t_{c}\right)$.

If we condition on an $\left(X,t_{c}\right)$ that encompasses multiple
$\left(Z_{i},t_{c}\right)$ (e.g., the outcome of $\varUpsilon$ at
$t$ was not obtained, or is unknown to us), we can still calculate
the probabilities of the $\left(Y_{i},t\right)$ as $P(Y_{i},t;X,t_{c})=\frac{1}{Tr(\mathbb{P}_{X}(t_{c})\mathcal{P}(t_{0}))}Tr\left(\mathcal{P}_{Z_{i}}(t_{c})\mathcal{P}(t_{0})\right)$.
However, it may not be possible to perform a measurement on $X$ at
$t_{c}$ to determine what the outcome would have been. This is because,
while the $\mathcal{P}_{Z_{i}}(t_{c})$ subspaces are disjoint, the
may not be orthogonal, and so may not commute (though that all so
commute with $\mathcal{P}_{X}(t_{c})$).

\subsection{The reference experiment}

We are now in a position to correctly calculate all probabilities
for our reference experiment. Take $I$ to be the set of possible
states of the experimental equipment when the first detector flashes
at time $t_{0}$, and $F_{\uparrow}$ to be set of possible states
of the experimental equipment when the upper half of the second detector
flashes at time $t_{1}$ (indicating that the particle spin is in
the x-direction). Given $I$ at time $t_{0}$, the probability of
$F_{\uparrow}$ at $t_{1}$ is: 

$P(F_{\uparrow};t_{1}|I;t_{0})=\frac{1}{Tr\left(\mathcal{P}_{I}(t_{0})\mathcal{P}(t_{s})\right)}Tr\left(\mathcal{P}_{F_{\uparrow}}(t_{1})\mathcal{P}_{I}(t_{0})\mathcal{P}(t_{s})\mathcal{P}_{I}(t_{0})\right)$

Where $t_{s}$ is the ``start time'', which is usually referred
to as $t_{0}$. 

In $\mathcal{P}_{I}(t_{0})$ (unlike $\mathbb{P}_{I}(t_{0})$) the
particle's spin points in the positive y-direction, and in $\mathcal{P}_{F_{\uparrow}}(t_{1})$
it points in the positive x-direction. So long as the Hamiltonian
used in the calculation faithfully reproduces the dynamics of the
SG measurement, $P(F_{\uparrow};t_{1}|I;t_{0})$ is then equal to
$\left|\left\langle s_{y}|s_{x}\right\rangle \right|^{2}$.

To calculate $P(I;t_{0}|F_{\uparrow};t_{1})$, first note that it
is explicitly assumed that the recording device in $F_{\uparrow}$
retains the information that $I$ was the case at $t_{0}$, and so
the probabilities are verifiable. This means that $\mathcal{P}_{I}(t_{0})\left(\mathcal{P}(t_{0})\mathcal{\mathbb{P}}_{F_{\uparrow}}(t_{1})\mathcal{P}(t_{0})\right)=\mathcal{P}(t_{0})\mathcal{\mathbb{P}}_{F_{\uparrow}}(t_{1})\mathcal{P}(t_{0})$.
It follows that $P(I;t_{0}|F_{\uparrow};t_{1})=1$. 

The modified Born rule therefore correctly yields all probabilities.

\subsection{Final remarks on verifiable probabilities}

The question of when probabilities are verifiable does not only arise
in quantum probabilities. In order to be able to verify probabilities
in any real world situation, something must keep a record of the events
that take place. The question is, however, particularly significant
it quantum probabilities, because cases where such a component is
lacking can lead to misunderstandings \& fallacies \footnote{A few years back, an interesting example of the difficulties that
arise when probabilities are not verifiable was published\citep[Frauchiger, et al,][]{Frauchiger}.
It described a variant of the Wigner's friend scenario. A number of
interlocking conditions were given, which together yielded contradictory
or paradoxical conclusions. However, the final measurement would destroy
the information present in the conditions, so the contradictory conclusions
about the final measurement based on the given conditions would not
be known once the measurement is made.}.

This does not mean that the Born rule is only applicable for verifiable
probabilities. It may be of interest to calculate a probability distribution,
even when there is no possibility of the measurement being performed.
An example of this was given in the last section, where the position
probability of a particle in a tracking detector, $\varrho(\vec{x},t)$,
was of interest even though the setup had no ability to perform the
extra position measurement. The restrictions of verifiable probabilities
do, however, apply when discussing experiments that actually are performed.

\section{More general probability sequences}

We may try to expand the Born rule by considering sequences of outcomes:
$P(Y_{2},t_{2};Y_{1},t_{1}|X,t_{c})=\frac{1}{Tr\left(\mathbb{P}_{X}(t_{c})\mathcal{P}(t_{0})\right)}Tr\left(\mathbb{P}_{Y_{2}}(t_{2})\mathbb{P}_{Y_{1}}(t_{1})\mathbb{P}_{X}(t_{c})\mathcal{P}(t_{0})\mathbb{P}_{X}(t_{c})\mathbb{P}_{Y_{1}}(t_{1})\right)$.
However, such calculations will only yield reliable results if the
sequence satisfies the restrictions of verifiability, in which case
it can be reduced to the previously given version of the Born rule.
Any other attempt at applying such a rule would be like asking, within
the context of a double slit slit experiment, what the probability
is of the particle going through one of the slits, and then ending
up in some final location, even though information about the slit
is not retained in the final state.

\section{A restriction on observers}

An observer may be defined as any physical system whose internal state
mirrors the state of the outside world. This is a fairly common circumstance.
The physical subspace helps to enable a quantum description of observers
by excluding state vectors in which, for example, a particle detector
in a state of detection is paired with a particle state that is not
within the detector.

Observations must satisfy a simple, though perhaps unexpected, self-consistency
condition. Let's say that a system2 subspace, $M$, is physically
possible, meaning $\left[\mathbb{P}_{M},\mathcal{P}\right]=0$ \&
$\mathcal{P}\mathbb{P}_{M}\neq0$. Let's also say that system1 contains
an observer that observes system2. Take some observer state, $O$,
that also satisfies: $\left[\mathbb{P}_{O},\mathcal{P}\right]=0$
\& $\mathcal{P}\mathbb{P}_{O}\neq0$. That is we only assume that
$O$ \& $M$ are physically possible, and that $O$ implies something
about the state of system2.

Can $O$ imply that system2 is in mixed state of $M$ \& $\neg M$?
If not, then system1 observers can never observe system2 in a mixed
state of $M$ \& $\neg M$.

It does appear that such an observation is impossible. Because $M$
describes system2 states, and $O$ describes system1 states, $\left[\mathbb{P}_{M},\mathbb{P}_{O}\right]=0$.
Applying the previous commutation relations, it also follows that
$\left[\mathbb{P}_{M},\mathcal{P}_{O}\right]=0$. Therefore, $O$
can not necessitate that system2 is in a mixed state of $M$ \& $\neg M$.

This result is due, in part, to how ``physically possible'' has
been defined; in some intuitive sense, the definition may be considered
too strict. Consider the portion of a physical subspace consisting
of outcomes of Stern-Gerlach experiments. (For obvious reasons, this
description will be somewhat simplified.) Call the system1 state corresponding
to measuring spin in the $\vec{s}$ direction $O(\vec{s})$. For spins
pointing in opposite directions, the $O(\vec{s})$ states correspond
to the same SG set-up with different outcomes. In all other cases,
the SG set-ups are different, because they measure spin along different
axes. Therefore $\left\langle O(\vec{s}_{1})|O(\vec{s}_{2})\right\rangle =\delta_{\vec{s}_{1},\vec{s}_{2}}$.
The physical subspace is then $\mathcal{P}=\sum_{\vec{s}}\left|O(\vec{s})\right\rangle \left|\vec{s}\right\rangle \left\langle \vec{s}\right|\left\langle O(\vec{s})\right|$.
The system1 $O(\vec{s})$ states are physically realizable, but the
system2 spin states are not. This is because the various spins are
not mutually orthogonal; $\left|\vec{s}\right\rangle $ \& $\left|-\vec{s}\right\rangle $
are possible outcomes of some SG experiment, but $\alpha\left|\vec{s}\right\rangle +\beta\left|-\vec{s}\right\rangle $
is also a possible outcome of one of the SG experiments.

And yet it can be reasonable to say that a particle is in some spin
state, $\left|\vec{s}\right\rangle $. In our physical subspace, the
ability to make a claim about the particle's spin state depends on
the state of system1: $\vec{s}$ is physically realizable if system1
is in state $O(\vec{s})$. This can be generalized to say, a subspace
$C$ is ``conditionally realizable'' if for some physically realizable
$R$, $\left[\mathbb{P}_{C},\mathcal{P}_{R}\right]=0$ and $\mathcal{P}_{R}\mathbb{P}_{C}\neq0$.
The closed system can then be in $C$, if $R$ holds true. (There's
nothing terribly profound here; it's just saying that $\mathbb{P}_{R}$
is physically realizable, and $\mathcal{P}_{R,C}$ is a non-vanishing
subspace of $\mathcal{P}_{R}$.)

None the less, the above result can impact the types of observations
that can be performed. For example, consider a scenario in which observer1
is in state $O_{1}$ if it observes a system in some state, and in
state $O_{2}$, if it observes the same system in different state.
Can a second observer then observe observer1 in a mixed state of $O_{1}$
\& $O_{2}$? As we would desire states of observation to be physically
possible, the answer is: No.

\section{The quantum probability calculus}

The full quantum probability calculus can now be described briefly
as: Every observable quantity is mapped to some Hermitian operator
acting in a Hilbert space, the operator's eigenvalues being the possible
values for the observable. From these, projection operators for sets
of observable values can be constructed. Within the Hilbert space,
there is a physical subspace that enables a description of observation
of these observables. With these various projection operators, all
probabilities of observable quantities can be calculated using the
Born rule, as given in Sec \ref{sec:The-Born-rule}.

This calculus is observationally complete. Any observable aspect of
our physical universe can, in principle, be predicted by it. 

A question of some significance, however, remains unanswered.

\section{Types}

We briefly turn to a basic question from linguistics, which may seem
quite removed from quantum theory, but has bearing on issues of interpretation.
The question is: How does a language's elements get their meaning?
Very commonly, words \& phrases do not obtain meaning from within
their language, they get it by mapping these words \& phrases onto
aspects of our world. For example, to define the phrase ``my coffee
mug'', I would not craft a sentence defining what is meant by ``my
coffee mug''; I would point to the mug and say ``that's my coffee
mug''. In this way, the phrase ``my coffee mug'' is associated
with an object in our world, and so obtains its meaning. Many, perhaps
most, elements in a language are defined is this manner. ``Nebraska'',
``water'', ``osteoporosis'', ``millisecond'', and so forth all
rely on a relationship being established between our world and the
language to obtain their meaning.

This also applies to the language employed by scientific theories.
Observable quantities such as ``time'', ``distance'', ``mass'',
etc., are all defined within our world by the means we use to measure
them. They are not defined by the theory. When a theory refers to
``time'', it is referring to the thing measured by clocks. A theory
can posit relationships among these observable quantities, but these
should not be considered definitions. The relations can fail, in which
case the theory is shown to be incorrect.

A theory's scientific meaning depends on this relationship being established
between elements of the theory, and our world. In quantum theory,
the relationship between our world and elements of the theory is captured
by the probability calculus. In it, the mapping from mathematical
objects onto aspects of our world does not only include observable
quantities like time, position, and momentum, it also includes ``probability''
and ``knowledge'' (as in, ``it is known that $X$ occurred at time
$t_{0}$''; this must mean the same thing within the theory as we
normally mean by ``it is known that $X$ occurred at time $t_{0}$'').
The meanings of these terms exist prior to the theory, and must mean
the same thing within the theory.

If we wish, we can start by viewing quantum theory as a purely mathematical
theory in which a few axioms are asserted, and from which results
can be deduced. We then endow various elements of the theory with
real world meaning - this operator refers to momentum, with the operator's
eigenvalues referring to values of momentum; this scalar refers to
time; this function calculates conditional probability; and so forth.
With that, the theory now makes claims about manner in which our world
operates. Any question about our world that can be posed in terms
of those aspects that are mapped onto the theory can be answered by
the theory. 

From this perspective, quantum theory lacks nothing. It is, as asserted
earlier, observationally complete. However, when viewed from another
perspective, the theory is surprisingly lacking.

One does not simply expect a theory to tell us facts about observable
quantities. One also expects it to describe the nature of the universe.
Yet on this subject, quantum theory says almost nothing. It does not
even describe the nature of its observable quantities. What is the
physical nature of momentum, or energy, or spin? We know how to measure
the quantities, and these measurements \& their probabilities are
correctly predicted by the theory. But the theory does not specify
what it is that is being measured.

There are, presumably, many different types of worlds that satisfy
quantum theory, and each of them fill in these blanks differently.
A multiplicity of different types of worlds that would look just like
ours, down to obeying the laws of quantum theory, but that operate
quite differently. Some well known examples of types of quantum worlds
are those described by Bohmian mechanics, those described by the many
worlds interpretation, and those described by physical collapse theories.
Each of these is (presumably) internally self-consistent, and satisfies
quantum theory, and so would look just like our world. But they are
of quite different natures.

Because all types of quantum worlds will satisfy quantum theory, there
is little reason to ask which one is correct. That is, which one corresponds
to how \emph{our} world actually works. At present, there is simply
no way to answer this question. Indeed, so long as quantum theory
continues to be found to be correct, there will never be a way to
answer this question. As matters stand, it is likely more useful to
accept that all types of quantum worlds are valid, and to consider
that to fully understand the theory it would be helpful to understand
all the various ways in which it can hold true.

That, however, is a project for another day. For now, we can content
ourselves with quantum theory's apparent observational completeness.
It may be interesting to note that textbook presentations of quantum
theory are \emph{not} observationally complete. As was shown in the
introduction, they can not fully describe real world experiments,
in particular they often can not capture the information that recordings
of experiments contain. That, however, turned out to be a minor affair.
With mild additions to the theory the issue is resolved, and the theory
now does appear to be observationally complete.

\end{document}